# Towards an optical FPGA - Programmable silicon photonic circuits

Xia Chen[*], Milan M. Milosevic, Antoine F.J. Runge, Xingshi Yu, Ali Z. Khokhar, Sakellaris Mailis, David J. Thomson, Anna C. Peacock, Shinichi Saito and Graham T. Reed

Optoelectronics Research Centre, University of Southampton, Southampton SO17 1BJ, United Kingdom

[*] Tel.: +44 23 8059 2697; Email: xia.chen@soton.ac.uk

**Abstract**

A novel technique is presented for realising programmable silicon photonic circuits. Once the proposed photonic circuit is programmed, its routing is retained without the need for additional power consumption. This technology enables a uniform multi-purpose design of photonic chips for a range of different applications and performance requirements, as it can be programmed for each specific application after chip fabrication. Therefore the cost per chip can be dramatically reduced because of the increase in production volume, and rapid prototyping of new photonic circuits is enabled. Essential building blocks for programmable circuits, erasable directional couplers (DCs) were designed and fabricated, utilising ion implanted waveguides. We demonstrate permanent switching between the drop port and through port of the DCs using a localised post-fabrication laser annealing process. Proof-of-principle demonstrators in the form of generic 1×4 and 2×2 programmable switching circuits were then fabricated and subsequently programmed, to define their function.

**Introduction**

Silicon photonics is regarded as one of the most promising technologies to realise large-scale integration of optical circuits [1, 2]. Due to its capability for high-density integration and low fabrication cost at high volume, silicon photonics is widely used as a high-speed and low-cost solution for short reach interconnects. It is also proposed that silicon photonics can be potentially used in a much wider range of applications, such as sensing and computing [3-7].

In this paper, we demonstrate a new technology for building programmable silicon photonics circuits, in which the system architecture can be permanently programmed on demand by a post-fabrication, localised laser annealing process. This is analogous to field-programmable gate arrays (FPGAs) for the electronic integrated circuits, where many logic blocks can be "wired" on demand into a desired configuration after device fabrication, in order to suit different applications. We have demonstrated high-efficiency erasable direction couplers (DCs) for silicon photonics circuits, and have used them to construct more sophisticated photonic switching circuits. By utilising this technology, after wafer fabrication, one single silicon photonics chip could be programmed to perform various functions.

The high cost caused by comparatively low production volume has been hindering the commercialisation and expansion of silicon photonics [8]. One major envisaged application of our proposed technology is to standardise the photonic circuit design for a vast variety of applications and therefore reduce the cost per chip. One multi-purpose chip can be designed and fabricated with many functional components, and later be programmed for specific applications. Such a photonic chip can then be fabricated in a much larger volume, and the cost for each chip can be therefore reduced by a few orders of magnitude, according to the JePPIX roadmap in 2016.

The proposed programmable photonic circuits will also significantly increase the functionality and open new possibilities for silicon photonics, allowing rapid prototyping of new functional circuits. There are fundamental differences between the proposed technology and the reconfigurable photonic circuits based on conventional optical switches and photonic logic gates [6, 9-18]. The conventional reconfigurable circuits require a constant source of optical or electrical power to perform and maintain each switching operation, and the power consumption will increase exponentially with the complexity of the photonic circuits. Furthermore, these techniques may not be robust enough for many applications in a harsh environments. Comparatively, our technique involves physically changing the routing of the photonic waveguide, requiring no additional control during operation. It is wideband and very robust to temperature change.

In this work we report integrated waveguides, which are formed by ion implantation damage, on the commonly used silicon-on-insulator (SOI) platform. By employing such implanted waveguides, erasable directional couplers have been demonstrated with high coupling efficiency, which form the basic building block of a programmable photonic circuit. As a proof-of-principle demonstration, we successfully realised 1×4 and 2×2 programmable switching circuits on a SOI wafer. Finally, by way of example, we present a vision for a multi-purpose transmitter chip design, which can be programmed after fabrication to support various commonly available transmission formats in the communication links.

**Erasable waveguides and directional couplers**

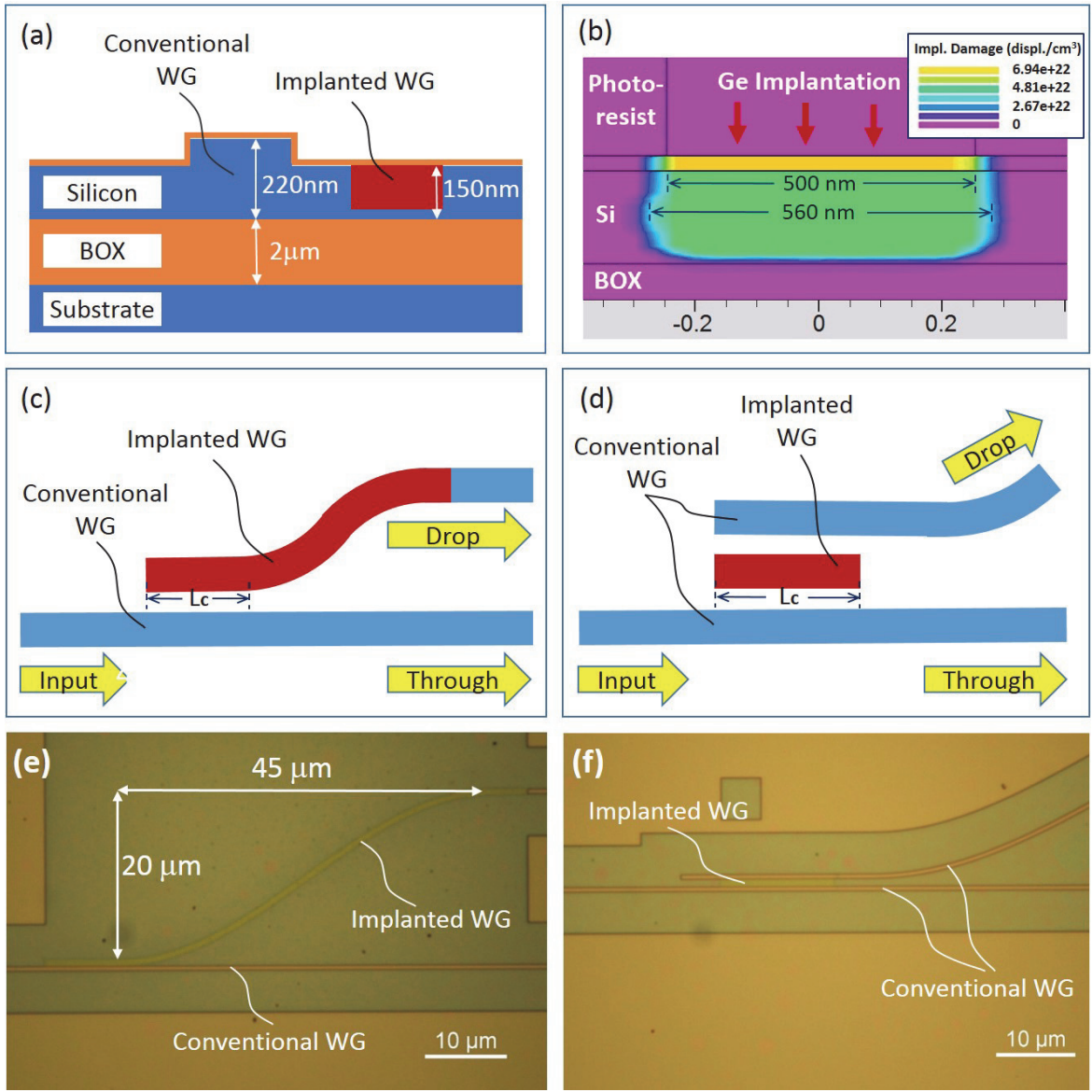

Figure 1. Structures of the implanted waveguides (WG) and DCs. **a**, Cross sectional view of the erasable DC with a conventional rib waveguide and an ion implanted waveguide. **b**, Simulation result for the Ge implanted process using the Silvaco software tool, the density of damage to the crystal lattice is shown in the figure. **c**, Illustration of the single-stage DC. **d**, Illustration of the two-stage DC. **e**. Optical microscope images of a fabricated a single-stage DC on a SOI wafer and; **f**. a two-stage DC.

Directional couplers (DCs) are one of most basic elements enabling a programmable circuit. Compared to other options for making switches, such as MZIs [19] and ring resonators [10], they can be much smaller in size, and footprint is a major consideration for measuring the cost in the semiconductor industry. Two types of DCs were design and fabricated, as shown in Fig.1c (single-stage DC) and Fig.1d (two-stages DC). A typical cross section of proposed devices is shown in Fig.1a. Implanted waveguides are successfully formed in the slab region of the conventional rib waveguide to couple light in/out, which can be erased by a laser annealing process (see methods).

Ion implantation induced index change was previously utilised to demonstrate wafer-scale testing [20, 21] and post-fabrication trimming applications[19, 22, 23]. However, for those work, the ion implantation was done only in preformed waveguides, in order to change the phase of optical signal. No integrated waveguide has been demonstrated in silicon photonics, which is formed solely by ion implantation process. Sufficient Ge ion implantation will create over 80% crystal lattice displacement in the silicon and a refractive index change of 0.5. The depth of the implanted waveguide is 140 nm into the silicon layer according to our calculation (Fig.1b). Optical microscope images of a typical fabricated single-stage DC and two-stage DC are shown in Fig.1e and Fig.1f, respectively.

Implanted waveguides with different widths were characterised. Experimental results are shown in Fig. 2. A linear dotted line was fitted for each device group measured and the function equations shown in the figure in the corresponding order. According to our measured results, the average propagation loss of the implanted waveguides are 28 dB/mm, 32.6 dB/mm, 32.4 dB/mm and 33.5 dB/mm for the 360 nm, 560 nm, 760 nm and 960 nm widths, respectively. This loss is high compared to conventional waveguides. However, only a short section of implanted waveguide (~10 μm) is needed for each switching device, corresponding to a propagation loss of only 0.3 dB. The loss may be further reduced by optimising the implantation dose or device design. The average transition loss for each conventional and implanted waveguide connection is about 0.9 dB. We believe that the propagation loss is mainly caused by absorption within the implanted silicon. The loss is therefore less for narrower waveguides as more optical power will propagate in the low-loss crystal silicon slab surrounding the implanted waveguide. The propagation loss increases with waveguide width as more light will be guided in the waveguide core of implanted silicon.

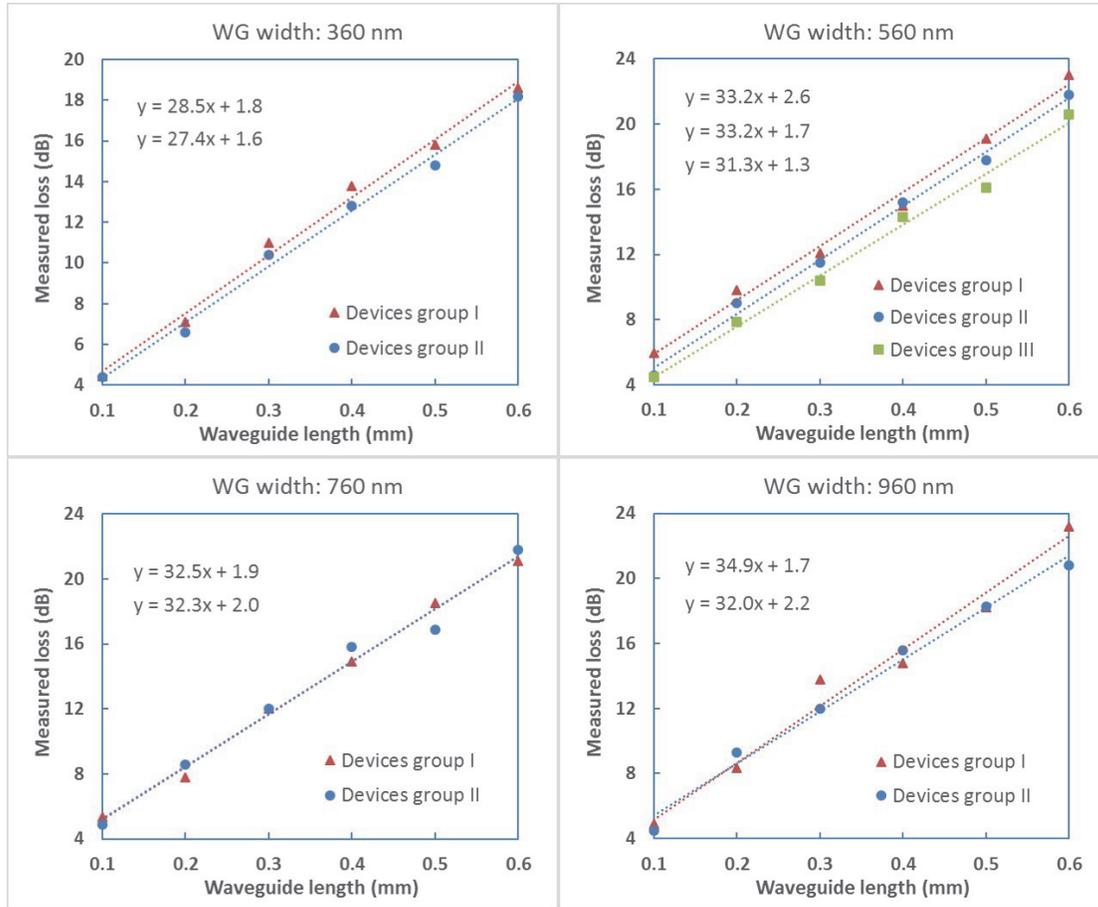

Figure 2. Measured propagation losses of the implanted waveguides. A linear dotted line was fitted for each device group measured. Each device group includes waveguides from the same silicon chip. The corresponding function equations are shown in the figure. The width of the implanted waveguide is: **a**, 360 nm, **b**, 560 nm, **c**, 760 nm, and **d**, 960 nm.

The measured results of single-stage DCs and two-stage DCs are plotted in Fig.3a and Fig.3b, respectively. Over 80% coupling efficiency was experimentally obtained for a coupling length of around 6 μm for the single-stage DCs. Calculated results are plotted as a reference. Over 90% coupling efficiency was obtained for coupling length around 12 μm for the two-stage DCs.

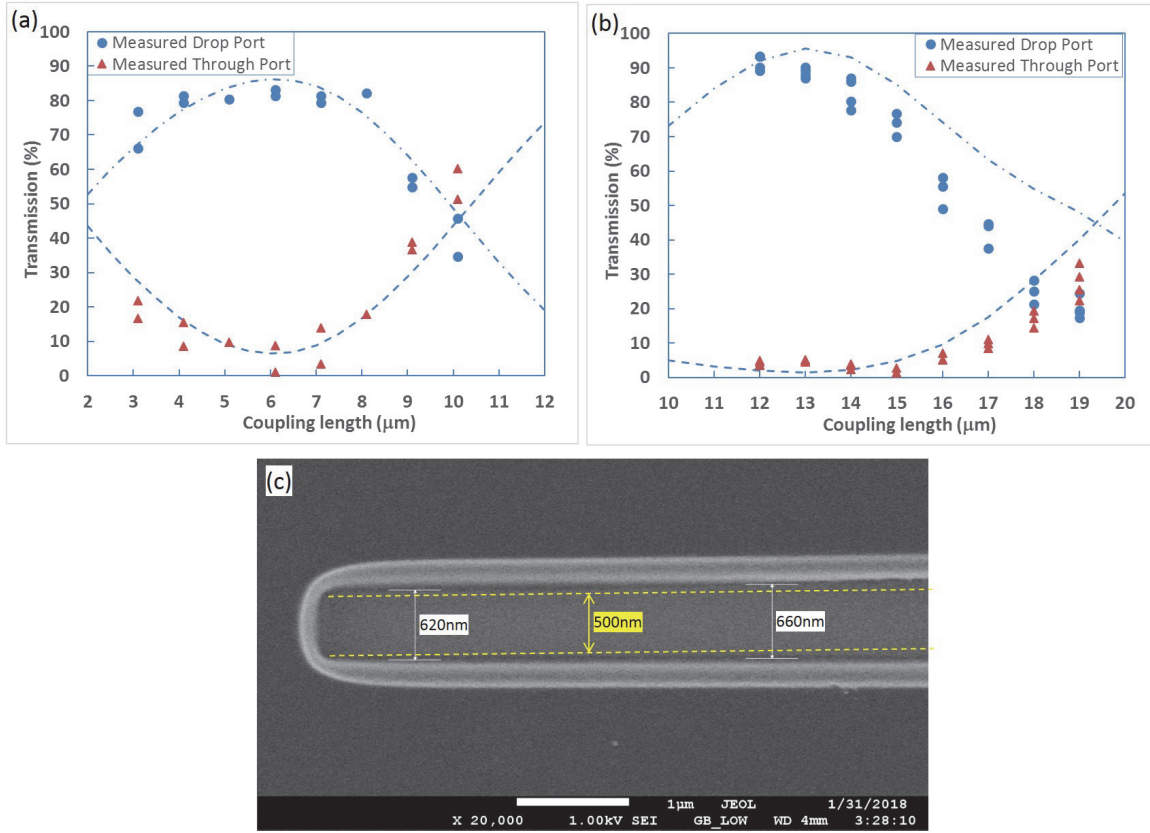

Figure 3. Measured optical transmission of the: **a**, single-stage DCs with various coupling lengths, and **b**, two-stage DCs with various coupling lengths. Simulated results (dotted line), with a step-index profile assumed for the implanted waveguides, are plotted as a reference. **c**, SEM image of a typical opening in the implantation mask after Ge ion implantation. The original designed width of the opening is 500nm (marked in yellow). However, the width was increased to around 620 - 660 nm after ion implantation.

Compared to our original simulation results, we found that the coupling lengths were shorter for both types of DCs. Such reduction of the coupling length is most likely caused by an increased overlap of the optical mode in the two coupled waveguides, which means that the actual optical mode in the implanted waveguide is wider than the optical mode used in our simulations. We noticed through SEM measurements that the E-beam resist, which was used as the implantation mask, shrank dring the implantation process as shown in Fig. 3c. A 500 nm wide opening developed for Ge implantation was increased to around 620 nm - 660 nm in width after implantation according to our measurement. This will lead to a wider implanted waveguide with a graded index profile, and therefore a reduced coupling length as we measured. The simulations results were adjusted accordingly, and plotted in Fig.3a and 3b. The shrinkage of the resist layer could be caused by the heat induced by the ion implantation process. Replacing the resist based implantation mask with a hard mask would eliminate this uncertainty in the fabrication process.

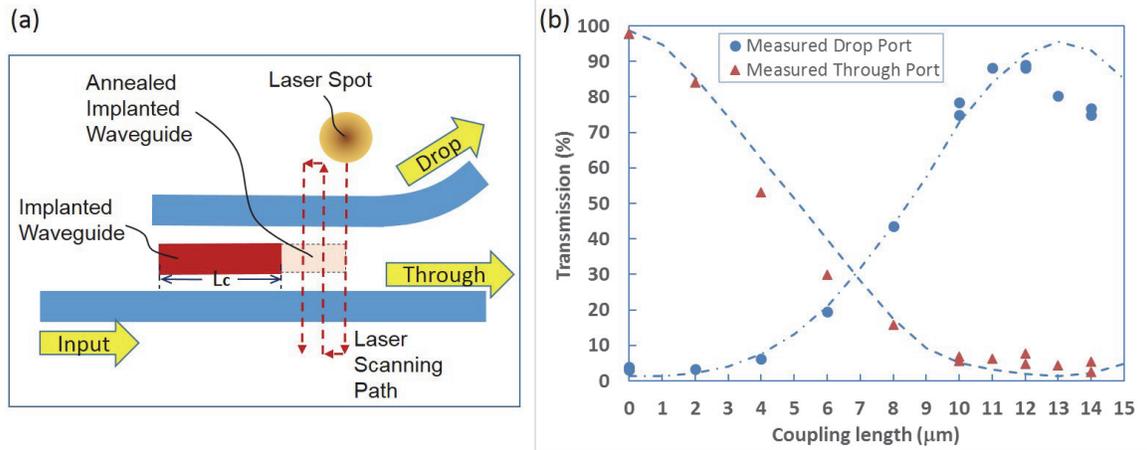

Figure 4. **a**, Illustration of the laser annealing process for the two-stage DCs. **b**, Measured results for the two-stage DCs with various coupling lengths after laser annealing. The original coupling length for the directional couplers used for this test are 20 μm. Simulation results (dotted line) were plotted as a reference.

A permanent optical switch is demonstrated by localised laser annealing of a two-stage DC, which is the most basic component for photonic programmable circuits. An illustration of the localized laser annealing process (see methods) is shown in Fig. 4a. The implanted waveguide was annealed by a focused laser spot in a step by step fashion, in order to maximize the coupling efficiency to the drop port with an optimum coupling length $L_c$. The raster scan step of the laser was 1 μm, to ensure uniform coverage of the area. Light propagating through the device will be mostly guided to the through port when the implanted waveguide is annealed completely ($L_c = 0$).

The experimental results of the laser annealing test for the two-stage DCs are plotted in Fig. 4b. For all the DCs measured in this experiment, the original coupling lengths (before annealing) were 20 μm. A few devices were characterised with various lengths of the implanted waveguide left unannealed (coupling length $L_c$). The measured results are in good agreement with our simulation results. A coupling efficiency of 89% (-0.5 dB) is achieved with a coupling length of approximately 11 or 12 μm, with an average residual transmission of 5% (-13 dB) to the through port. When the implanted waveguide is fully annealed, there is approximately 3% (-15 dB) cross coupling to the drop port, with 97% (-0.13 dB) transmission the through port.

**Demonstration of the photonic programmable circuits**

Based on the permanent setting of the optical coupling described previously, we have demonstrated 1×4 and 2×2 switching circuits as proof-of-principle demonstrations of a permanent programmable circuit implemented with our proposed technology. The microscopic images of the switching circuits are shown in Fig.5a and 5b. The 1×4 switching circuit comprises three two-stage DCs in series. The drop ports of those DCs lead to outputs P1, P2, and P3, respectively. The through port leads to output P4. After device fabrication, the optical signal can be permanently programmed to output at any of the four output ports, depending on the specific requirements arising from the application. For the fabricated device, the coupling length was designed to be 13 µm for all of the DCs. This coupling length is not yet optimised to the best efficiency to the drop port. Therefore, we need an additional step in the programing algorithm to anneal the coupling length to 11.5 µm to maximum the coupling efficiency to the corresponding drop port, as suggested by the results from Fig. 4. Clearly, if the optimal coupling length was to be directly fabricated in the future, then this initial programming step can be avoided.

The algorithm for programming the output to port P1 is simply annealing the first DC to its maximum coupling efficiency to the drop port. The measured results for all the four output ports are shown in Fig. 5c. The transmission is about -0.5 dB for P1, which matches well with our previously results for the discrete two-stage DC devices. The extinction ratio against the transmission to P2 is 14.4 dB at 1550 nm wavelength as plotted in the figure. For programming the output to port P2, we annealed the whole implanted waveguide for the first DC, and then annealed the coupling length to 11.5 µm for the second DC to optimise its coupling efficiency to P2. As shown in Fig. 5d, the transmission to P2 is approximately -0.8 dB, which includes a -0.13 dB loss at the first DC and a -0.5 dB loss while coupling to the drop port at the second DC. There is a 0.17 dB discrepancy between the expected loss values and the measured result at 1550 nm, which could be attributed to the measurement uncertainty and the imperfections associated with the fabrication process.

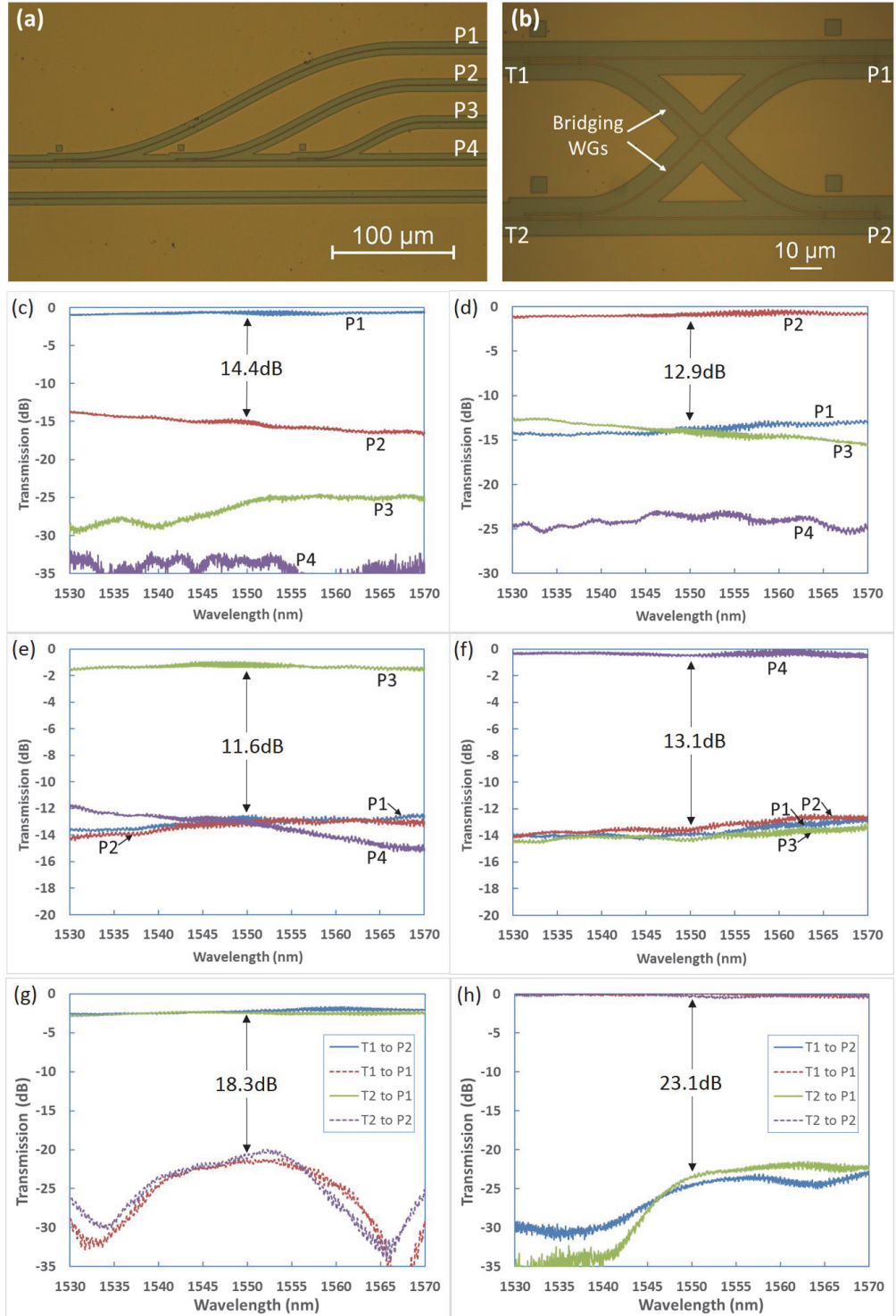

Figure 5. **a**. Optical microscope images of a 1×4 programmable photonic switching circuit and **b**. a 2×2 photonic switching circuit. Measurement results for the photonic switching circuits. The 1×4 photonic switching circuit was programmed by laser annealing to produce an output at one of the four ports sequentially (**e, d, e** and **f**). **c**. Measured results when P1 is set to be the output port. **d**. Measured results when P2 is set to be the output port. **e**. Measured results when P3 is set to be the output port. **f**. Measured results when P4 is set to be the output port. The 2×2 switching circuit was programmed to operate in two modes by laser annealing: **g**. cross-coupling modes or, **h**. through-coupling mode.

Similarly, for programing the output to port P3, we annealed the whole implanted waveguide for the first (P1) and the second DC (P2), and then annealed the coupling length to 11.5 μm for the third DC (P3). As shown in Fig. 5e, the transmission to P3 is -1.1 dB in average, which includes a -0.26 dB loss from the two fully annealed DCs and a -0.5 dB loss while coupling to the drop port at the third DC. There is again a 0.34 dB discrepancy between the expected loss values and the measured result at 1550nm, which could also be attributed to the measurement uncertainty and the imperfections associated with the fabrication process. The measured extinction is 11.6 dB, which is the lowest number among the four switching states. This is mainly because of the relatively high transmission loss. When all the three DCs were fully annealed, the optical signal is guided through the switching circuit to P4. The measured results in this case are plotted in Fig. 5f. The transmission is about -0.5 dB for P4 in this case, which is because of the cross coupling loss of the three fully annealed DCs. We measured a 13.1 dB extinction ratio for this state.

The 2×2 switching circuit is comprised of a 2×2 array of two-stage DCs arranged in a pattern as shown in Fig. 5b to enable cross coupling between two signal paths. There are two signal inputs T1, T2 and two outputs P1, P2. Although a simple 2×2 switch can be made using the two outputs and inputs of a single DC, our design with the 2×2 array of DCs can offer a much lower crosstalk [24] and more functionality. After device fabrication, the optical signal can be permanently programmed into two operating modes: cross-coupling mode and through-coupling mode, depending on the requirements arising from the applications.

When all the DCs are operating at the optimum coupling point to the drop port (with Lc = 11. 5 μm), the 2×2 switching circuit is operating in cross-coupling mode. The optical signal inserted in T1 will be coupled to a bridging waveguide (conventional shallow-etched waveguide) at the first DC, and then coupled over to P2 at the next DC. Similarly, the optical signal inserted in T2 will be coupled to P1. The measured results are shown in Fig. 5g. The transmission losses of T1-P2 and T2-P1 channel are about -2.3 dB on average, which includes approximately -0.5 dB insertion loss at each two-stage DCs in addition to the insertion loss of the waveguide cross in the middle of the bridging waveguide. The loss of the waveguide cross was characterised with separate testing structures. The average loss is approximately -0.6 dB. However, this loss figure can be potentially improved to only -0.16 dB with improved design [25-27]. There is -20.6 dB cross talk at 1550 nm according to the measurement results for both transmission channels, corresponding to an 18.3 dB extinction ratio (Fig. 5g).

When all the implanted waveguides of the two-stage DCs are annealed by the laser, the 2×2 switching circuit is working in through-coupling mode, when the light signals can go straight through the two DCs at each side from T1 to P1, or from T2 to P2. The measured results are shown in Fig. 5h. The transmission of both channels (T1-P1 and T2-P2) is around -0.2 dB on average, which is also expected due to the insertion loss of the two fully annealed DCs. The extinction ratio in this through-coupling mode is approximately 23.1 dB.

**System integration**

There are currently two major markets for silicon photonic devices: broadband optical communication networks and high-performance computing systems [28], and in each of them there are even more subcategories that pose different requirements on the transceivers employed in the data links, for example, in terms of bandwidth, link distance, power budget, cost per bit, etc. In order to meet those requirements, different types of silicon photonic transceivers are available, such as 100G PSM4 (Parallel Single Mode fibre 4-lane) optical transceiver, 4×25G WDM4 (4 channels Wavelength-Division Multiplexing) transceiver, or the more advanced QAM16 (16-ary Quadrature Amplitude Modulation) [29]. The architecture of the silicon photonic transmitter chip that is required for the transceivers in each of these cases are all different. The production volume of each silicon photonic chip is therefore hindered by this market segmentation. However, with the technology demonstrated in this paper, it is now possible to make one single silicon photonic chip design (as shown in Fig.6a) in large volume, and then programme it to suit various applications, such as PSM4 (Fig.6b), WDM4 (Fig.6c) and QAM (Fig.6d).

For the PSM4 transmitter in Fig.6b, we just erase all of the DCs. The multiplexer, power slitter and combiner and 90° phase shifter will be disabled. If we then keep the four DCs leading to the multiplexer, and keep the one DCs coupling back to the output port as shown in Fig.6c, the photonic chip will transmit signals in a WDM4 format. If we keep the beam splitter and combiner instead shown in Fig.6d, the input light will be guided into two modulators with a 90° phase difference. QAM signals can be generated with this architecture.

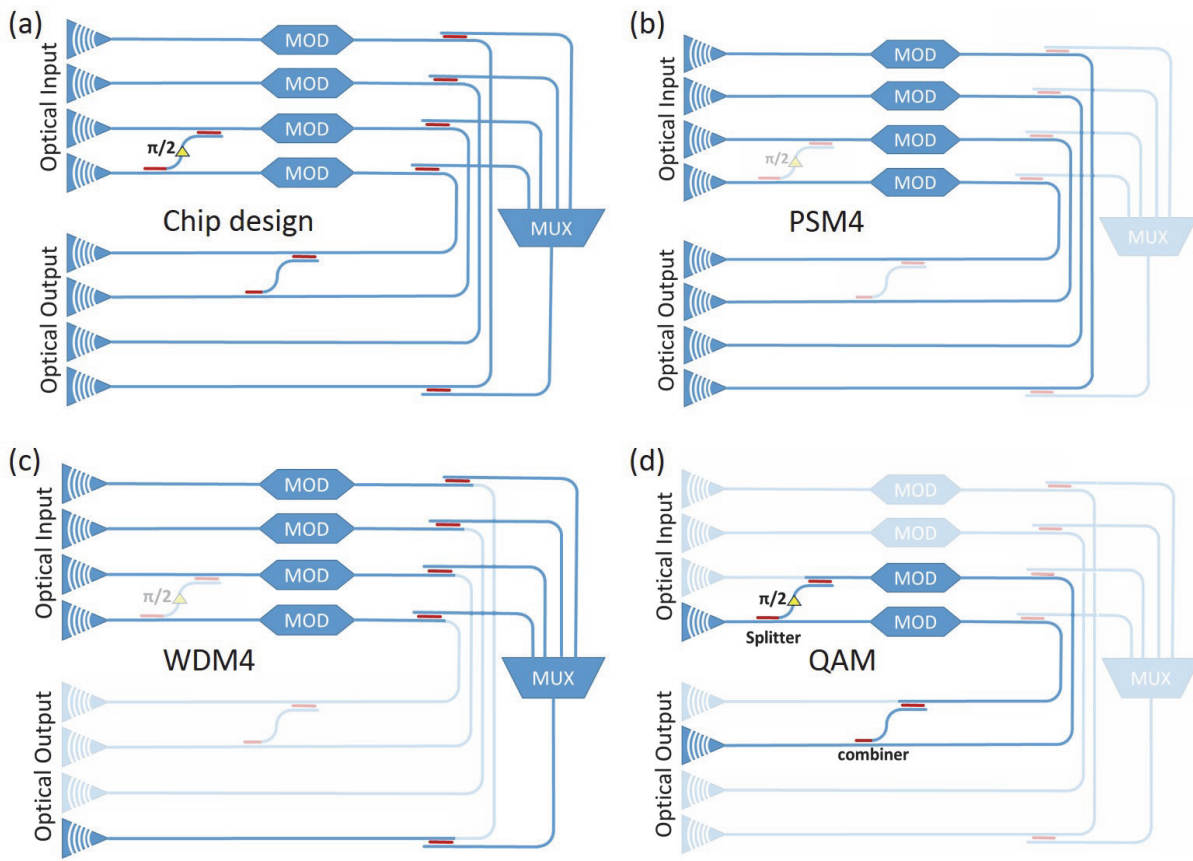

Figure 6. Illustration of a single photonic transmitter circuit that can be programmed for various transmission formats. **a**. The original chip design. **b**. Programmed transmitter chip for PSM4 transmission. **c**. Programmed transmitter chip for WDM4 transmission. **d**. Programmed transmitter chip for generating QAM signals. (The functional waveguides and devices are darker in colour. MOD: optical modulator, MUX: optical multiplexer.)

More generic photonic programmable circuits can also be built based on more advanced algorithms [13-16]. For most applications, energy consumption is one the key metrics for photonic devices. Previously demonstrated integrated linear photonic circuits require constant power to maintain each switching component, and the power consumption will grow exponentially with the number of inputs and outputs the linear circuit has. Comparatively, the technology proposed here requires no additional power after it has been programmed into the target configuration, making it a potentially key technology that enables low cost and low power consumption silicon photonic devices.

**Conclusion**

The essential building blocks, single-stage and two-stage DCs were designed and fabricated incorporating the implanted waveguides. By using waveguides formed by ion implantation damage, the output light can be switched permanently between the drop port and through port by a localised laser annealing process. About 89% (-0.5 dB) and 97% (-0.13 dB) coupling efficiency was experimentally demonstrated while coupling to the drop port and through port, respectively, for a two-stage DC. Further reduction of the transmission loss of the DCs is possible if we can reduce the coupling length and implantation dose, in order to reduce the optical absorption due to the defect states, whilst keeping the high coupling efficiency. 1×4 and 2×2 switching circuits were fabricated and characterized as proof-of-principle demonstrations. The circuits have been successfully programed using laser annealing to operate in all the possible architectures. It is possible to achieve a transmission of over -1 dB for all the channels with an improved waveguide propagation and crossing loss. We also showed a vision of one sample application where one single photonic chip can be programmed into a transmitter chip for PSM4, WSM4 or QAM transmission formats. Because of the low cost and low power consumption, such platforms could be instrumental in the commercial success of the silicon photonics industry.

**Methods**

Devices design and simulations

A variational FDTD solver from Lumerical MODE solutions was employed to calculate the performance of the DC designs at a wavelength of 1550 nm. A 10 nm mesh size around the core waveguide area was used in our calculations. The refractive index of the core region of the implanted waveguide is approximately 3.96 yielding a Δn of 0.48, which is in good agreement with the expected Δn from other literature at a wavelength of 1.55 μm. The imaginary part of the index of the Ge implanted Si is calculated to be 0.00085, based on the 33 dB/mm propagation loss as experimentally measured. The profile of the implantation damage density in silicon is calculated by Silvaco software.

A single-stage DC consists of a 500 nm wide conventional input waveguide, a straight implanted waveguide for light coupling, an S-bend made of an implanted waveguide and a conventional output waveguide (Fig.1c). The length of the straight implanted waveguide is defined as the coupling length $L_c$. The S-bend is in the shape of a sinusoidal curve of 45 μm in length and 20 μm in displacement as shown in Fig.1e. Approximately 93% (-0.32 dB) coupling efficiency should be achieved according to our simulation results for a coupling length of 8 μm and an ion implanted width of 540 nm, and 200 nm gap between the conventional and implanted waveguide. A two-stage DC, as shown in Fig.1f, consists of a conventional input waveguide, a straight ion implanted waveguide for coupling the light and transferring it to another conventional waveguide coupling to the drop port (also illustrated in Fig. 1d). The length of the implanted waveguide is defined as the coupling length $L_c$, as shown in Fig. 1d. The gap between the input waveguide and the implanted waveguide is the same as the gap between the implanted waveguide and the output waveguide to the drop port. Our simulations suggest a coupling efficiency of over 95% (-0.22 dB) is achievable for such two-stage coupling structures with a 200 nm gap, a 15 μm coupling length and a width of 560 nm for the implanted waveguide.

In our simulation, we defined a step-index profile for the implanted waveguide, where the index is 3.96 in the implanted region and then changed to crystal silicon (3.48 for 1550 nm wavelength) outside the implanted region. The width of the implanted region was defined by considering the region with over 80% crystal lattice displacement, which is 30 nm wider than actual opening of the implantation mask at each edge according to our calculations. In addition, there is a shrinkage of the implantation mask made of photo resist during ion implantation, as mention previously in this paper. The implanted waveguide is actually wider (70 nm added for each edge) than our original design. Our simulation result was corrected accordingly. For example, we increased the width of the step-index implanted waveguide from 540 nm to 680 nm (70 nm added each side) for the single-stage DCs, and slightly reduced the refractive index of waveguide core from 3.96 to 3.92, in order to approximate the graded-index profile actually created by the Ge implantation because of the shrinking mask. This problem can be solved using a hard mask for Ge ion implantation.

Devices fabrication and characterization

All DCs were fabricated on an SOI platform, with a 220 nm top silicon layer and a 2 μm thick buried oxide (BOX) layer (Fig.1a). Conventional 500 nm wide SOI rib waveguides were formed by a 70 nm partial etch (leaving a 150 nm silicon slab layer) to support only single-mode propagation, which is typically used for silicon photonic circuits. The conventional waveguides were fabricated by electron beam lithography and a plasma etching process. A 25 nm thick oxide layer was then deposited by Plasma-Enhanced Chemical Deposition (PECVD) as a protective layer. After this fabrication step, a 400 nm thick photoresist layer was deposited as a mask layer for Ge ion implantation. An ion energy and fluence of 130KeV and $1\times10^{15}$ ions/cm$^2$ were used, respectively. The implanted waveguides were then formed in the slab layer next to the conventional waveguide.

The loss of the implanted waveguides was characterised by measuring the output power of implanted waveguides with different lengths. It is worth mentioning that the implanted waveguide width is defined by the width of the region with over 80% lattice disorder, which is 60 nm wider than the actual opening of the implantation mask according to our calculations, as illustrated in Fig.1d. For testing, conventional surface grating couplers and waveguides were used to access the implanted waveguides. A simple butt-coupling structure is used to connect the conventional waveguide and the implanted waveguide. All the measured optical transmissions were normalised to the fibre-to-fibre optical loss of a reference waveguide in each case, which consists of only the grating coupler at the input and output, and a conventional waveguide with the same length as that used for coupling to the implanted waveguides or the photonics devices. For testing purposes, only the fundamental TE mode is coupled to the waveguides. When characterizing the single-stage DCs, the additional propagation loss in the S-bend and the transition loss to conventional waveguide were deducted.

Laser annealing

A continuous wave (CW) Argon ion laser was used for the annealing process. The laser operates in free running mode. The output is a mix of all lasing lines in the argon ion spectrum, dominated by 488 nm and 514.5 in almost equal parts. A CW laser based annealing has been shown to produce larger silicon crystal grain sizes with fewer defects than those obtained by pulsed laser source systems, resulting in a qualitative improvement in the crystallization of the amorphous silicon for enhanced material quality [30]. This has been attributed to the lower cooling rate after laser irradiation from continuous wave sources. The laser beam was guided to the surface of the silicon chip in a controlled and precise manner. A half wave plate and a polarization beam cube splitter were used to adjust the laser power. The laser beam was focused onto the chip surface using a microscope objective lens. A 20× objective lens was used, which produced a laser beam of approximately 2 μm in diameter. The sample was scanned under the laser beam using a set of linear micro-precision stages while a pellicle beam splitter and a CCD camera were used for imaging and control. The scanning speed was 10 μm/s.

**Acknowledgements**

This work was funded by EPSRC project under the "Silicon Photonics for Future Systems", "Electronic-Photonic convergence", "Laser-Engineered Silicon" and "CORNERSTONE" projects. Reed is a Royal

Society Wolfson Research Merit Award holder. He is grateful to the Wolfson Foundation and the Royal Society for funding of the award. Thomson acknowledges funding from the Royal Society for his University Research Fellowship.

**Author contributions**

Chen proposed the idea, fabricated the devices, carried out the characterisation of all sample, performed data analysis and prepared the manuscript. Milosevic and Thomson contributed to the device design and experimental study. Runge, Mailis and Peacock contributed to the laser annealing process. Yu contributed to the characterisation of some devices. Khokhar provided fabrication support. Saito and Reed provided the overall supervision and technical leadership. All authors reviewed the manuscript.